\documentstyle[epsfig,prl,twocolumn,aps,floats]{revtex} 
\begin{document}
\def\thefootnote{\fnsymbol{footnote}}
\topmargin 0cm 
\newcommand{\be}[1]{\begin{equation} \label{(#1)}}
\newcommand{\ee}{\end{equation}}
\newcommand{\ba}[1]{\begin{eqnarray} \label{(#1)}}
\newcommand{\ea}{\end{eqnarray}}
\newcommand{\nn}{\nonumber}
\newcommand{\rf}[1]{(\ref{(#1)})}
\def \lsim {\mbox{${}^< \hspace*{-7pt} _\sim$}}
\def \gsim {\mbox{${}^> \hspace*{-7pt} _\sim$}}
%%%%%%%%%%%%%%%%%%%%%%%%%%
\def\VEV#1{\left\langle #1\right\rangle}
\def\fig#1{{Fig. \ref{#1}}}
\def\eq#1{{eq. (\ref{#1})}}
\def\Eq#1{{Eq. (\ref{#1})}}
%%%%%%%%%%%%%%%%%%%%%
\def\ne{\hbox{$\nu_e$ }}
\def\nm{\hbox{$\nu_\mu$ }}
\def\nt{\hbox{$\nu_\tau$ }}
\def\nr{\hbox{$\nu_R$ }}
\def\O{\hbox{$\cal O$ }}
\def\L{\hbox{$\cal L$ }}
%%%%%%%%%%%%%%%%%%%%%%%%%%%
\def\21{$SU(2) \otimes U(1) $}
\def\npb#1#2#3{{\it Nucl.\ Phys.\ }{\bf B #1} (#2) #3}
\def\plb#1#2#3{{\it Phys.\ Lett.\ }{\bf B #1} (#2) #3}
\def\prd#1#2#3{{\it Phys.\ Rev.\ }{\bf D #1} (#2) #3}
\def\prep#1#2#3{{\it Phys.\ Rep.\ }{\bf #1} (#2) #3}
\def\prl#1#2#3{{\it Phys.\ Rev.\ Lett.\ }{\bf #1} (#2) #3}
\def\mpla#1#2#3{{\it Mod.\ Phys.\ Lett.\ }{\bf A #1} (#2) #3}
\def\sjnp#1#2#3{{\it Sov.\ J.\ Nucl.\ Phys.\ }{\bf #1} (#2) #3}
\def\jetpl#1#2#3{{\it Sov.\ Phys.\ JETP Lett.\ }{\bf #1} (#2) #3}
\def\rnc#1#2#3{{\it Riv. Nuovo Cimento }{\bf #1} (#2) #3}
\def\yf#1#2#3{{\it Yad.\ Fiz.\ }{\bf #1} (#2) #3}
\def\hepph#1{{\tt hep-ph/#1}}
\def\pc{private communication}
\def\opc{\hbox{{\sl op. cit.} }}
\def\ip{in preparation}
\twocolumn[\hsize\textwidth\columnwidth\hsize\csname @twocolumnfalse\endcsname
\rightline{hep-ph/0011150}
\rightline{FERMILAB-Pub-00/250-T}
\rightline{IFIC/00-68, IFT-00/26}
\title{ Neutrino Unification }
\author{P. H. Chankowski${}^1$, A. Ioannisian${}^{2,3}$, S. Pokorski${}^{1,4}$
  and J.\ W.\ F.\ Valle$^{2}$}
  \address{$^1$ Univ. of Warsaw Inst. Physics Ul. Ho\.za 69, 00-681,
    Warsaw, Poland}
\address{$^2$ Instituto de F\'{\i}sica Corpuscular -- C.S.I.C. -- Univ. of Val\`encia  \\
     Edificio Institutos de Paterna -- Apartado de Correos 2085 -
    46071  Val\`encia, Spain}
\address{$^3$ On leave from Yerevan Physics Institute, Alikhanyan Br.2,
  Yerevan, 375036, Armenia}
\address{$^4$ Frontier Fellow at Fermilab, P.O. Box 500, Batavia IL 60510, USA}
\date{\today} 
\maketitle
\begin{abstract} 
  Present neutrino data are consistent with neutrino masses arising
  from a common seed at some ``neutrino unification'' scale $M_X$.
  Such a simple theoretical ansatz naturally leads to quasi-degenerate
  neutrinos that could lie in the electron-volt range with neutrino
  mass splittings induced by renormalization effects associated with
  supersymmetric thresholds. In such a scheme the leptonic analogue of
  the Cabibbo angle $\theta_{\odot}$ describing solar neutrino
  oscillations is nearly maximal.  Its exact value is correlated with
  the smallness of $\theta_{reactor}$.  These features agree both with
  latest data on the solar neutrino spectra and with the reactor
  neutrino data.  The two leading mass-eigenstate neutrinos present in
  \ne form a pseudo-Dirac neutrino, avoiding conflict with
  neutrinoless double beta decay.
\end{abstract}
\vskip2pc]

%\section{Introduction}

The standard model (SM) and its minimal supersymmetric extension
(MSSM) fail only in accounting for the solar and atmospheric neutrino
data \cite{solar,skatm00,atmos} which strongly indicate the need for
neutrino conversions, as would arise from neutrino-mass-induced
oscillations \cite{latestglobalanalysis}.
It is thus reasonable to investigate simple theoretically motivated
ansatz that can account for these observations.

Inspired by the idea of unification of fundamental
interactions~\cite{unif} here we propose the idea of neutrino
unification: namely that the neutrino mass and mixings observed at low
energies take a very simple form at some high energy scale $M_X$.
Thus, we add to the basic Lagrangean the dimension--five
operator~\cite{Weinberg:1978uk}
\begin{eqnarray}
\frac{\lambda_0 \delta_{ab} }{M_X} (\phi\ell_a)(\ell_b \phi)~+~H.~c.
\label{eqn:ans}
\end{eqnarray}
where $\ell_a$ denote the three lepton doublets and $\phi$ is the
standard Higgs doublet. As a working hypothesis, we assume this
operator to be characterized at the scale $M_X$, by a single real
dimension--less parameter $\lambda_0$.  The breaking of the
electroweak symmetry due to a non-zero vacuum expectation value (VEV)
$\VEV{\phi}$ will generate, in addition to the known SM masses, the
seesaw-type neutrino mass operator $M_\nu$
\be{mass} 
M_\nu =  \frac{\VEV{\phi}^2}{M_X}  \lambda_0
\ee 
For $M_X \sim M_{Planck}$ neutrino masses would be too small to
account for the atmospheric neutrino anomaly. Thus we will adopt an
intermediate value for the $M_X$ scale.  In the basis in which the
charged lepton Yukawa couplings are diagonal (weak basis), $\lambda_0
\delta_{ij}$ gets transformed into
\be{l5weak} 
\Lambda = \lambda_0 \Omega^T \Omega
\ee 
where $\Omega$ is an arbitrary unitary matrix.  After the electroweak
symmetry breaking we can rotate back the neutrino fields,
$\nu=\Omega^{-1}\nu^\prime$, and in this basis (the charged lepton
mass matrix is still diagonal) we get
\begin{eqnarray}
{\cal L}\supset{\lambda_0v^2\over M_X}\nu^\prime\nu^\prime
-{g\over\sqrt2}\overline{e_L}\bar\sigma^\mu W^-_\mu U\nu^\primeñ+~H.~c.
\label{eqn:lagr}
\end{eqnarray}
where $U=\Omega^\dagger$ and the fields $\nu^\prime$ are the neutrino
mass eigenstates (in this paper we number the eigenvalues so that
$\Delta m^2_{21}=m^2_2-m^2_1\equiv\Delta m^2_\odot$ corresponds to the
solar neutrino oscillations and $\Delta
m^2_{32}=m^2_3-m^2_2\equiv\Delta m^2_{atm}$ corresponds to the
atmospheric neutrino oscillations).  In general, the mixing matrix $U$
is characterized by three mixing angles and three $CP$ violating
phases, one Dirac plus two Majorana-type phases
\cite{Schechter:1980gr,Schechter:1981gk,BIL} and can be written as
the product of $V_{CKM}$ by
\begin{eqnarray}
\left(\matrix{1&0&0\cr0&e^{i\alpha_1}&0\cr0&0&e^{i\alpha_2}}
\right)\label{eqn:matrix}
\end{eqnarray}
where we have chosen to parametrize $V_{CKM}$ exactly as the
Cabibbo-Kobayashi-Maskawa matrix describing quarks, with the familiar
three angles and one phase $\delta$.  It is clear that with our ansatz
(\ref{eqn:ans}) we have an additional freedom of performing any
arbitrary real $3\times3$ rotation of the neutrinos. The general form
of such a restricted matrix $U$ can be derived by applying a
convenient parametrization \cite{Schechter:1980gr} of a general
unitary matrix to the matrix $M\equiv\Omega^T\Omega$ and by imposing
the symmetry condition $M=M^T$. Using the freedom of re-phasing the
charged leptons, one can show that the $U$ matrix cannot be eliminated
by redefining the neutrinos.
Thus, despite the mass degeneracy amongst the neutrinos, they exhibit
non-trivial mixing effects~\cite{Valle:1987gv}. In contrast to the
case investigated in ref.~\cite{Valle:1987gv} where neutrinos were
massless but non-orthonormal, here it is the fact that the mass
eigenvalues have different phases which prevents one from rotating
away the mixing amongst the neutrinos. Nevertheless, the structure of
the mixing in our case is more restricted than the general case, since
only two angles and one phase remain physical.
The above construction has the virtue of explicitly listing the
relevant mixing parameters for exactly degenerate Majorana neutrinos.
Once the neutrino mass degeneracy is lifted by quantum corrections the
mixing effects require the full matrix $U$ with, however, a
non-trivial relationship amongst the three mixing angles. 

Non-zero values of the Dirac phases mean $CP$ non-conservation, while
Majorana phases of $\pi/2$ indicate different $CP$ parities of the
neutrino mass eigenstates and do not imply $CP$ violation
\cite{Schechter:1981hw}.  Thus in our case $CP$ is conserved if the
phase $\delta=0$ mod $\pi$ and $\alpha_1 , \: \alpha_2 =0$ mod $\pi/2$.
Our main conclusions do not depend on whether or not $CP$ is conserved
in the neutrino sector and from now on we will assume it is conserved.
Different $CP$ parities of neutrino mass eigenstates can then be
accounted for by different signs of neutrino masses (which for
$\alpha_1=\pi/2$ and/or $\alpha_2=\pi/2$ can be achieved by simply
absorbing these phases in the neutrino fields).

Now we turn to the renormalization effects.  One-loop renormalization
group equations (RGEs) for the $\Lambda$ coefficients characterizing
the dimension-5 non-renormalizable terms can easily be written down
both for the SM and MSSM~\cite{RGE}.  The RGEs take a particularly
useful form when written directly for mass eigenvalues and for the
elements of the mixing matrix \cite{CHKRPO}. The flavour independent
corrections to the $\Lambda = \O(1)$ coefficients are irrelevant for
our discussion. The flavour-dependent corrections are due to lepton
Yukawa couplings and, to a good approximation, determined by the
$\tau$ Yukawa coupling. Using e.g.  the results of ref. \cite{CHKRPO}
one can easily verify that, with the ansatz (\ref{eqn:ans}), such
effects cannot explain $|\Delta m^2_{32}|\gg|\Delta m^2_{21}|$ with
phenomenologically acceptable mixing angles.

It is remarkable that supersymmetry can induce flavour-dependent
threshold corrections associated with slepton mass splittling
\cite{CHUNPO} which can dominate over the $\tau$ Yukawa corrections.
We show in this paper that such corrections can lead to the desirable
low energy pattern with the ansatz (1). This is possible provided the
soft supersymmetry breaking scalar mass terms deviate sufficiently
from universality.

The quantum corrections to our ansatz (\ref{eqn:ans})
in the basis defined by eq. (\ref{eqn:lagr}) are given by \cite{CHUNPO}
\begin{eqnarray}
m_\nu^{ab}=m_a\delta_{ab}+m_a(U^TIU^\ast)_{ab}+m_b(U^\dagger I U)_{ab}
\label{eqn:massmatcor}
\end{eqnarray}
where in our case all $|m_a|=m$ up to a common flavour-independent
renormalization $\O(1)$ factor.  Here the indices $a,b$ refer to our
starting basis used in \eq{eqn:ans}.  The correction $I$ (calculated
in the electroweak charged lepton mass eigenstate basis) consists of
two parts
\begin{eqnarray}
I=I^{RG}+I^{TH}
\end{eqnarray}
where $I^{RG}$ is the renormalization group correction
\cite{Ellis:1999my} and $I^{TH}$ are the electroweak scale threshold
corrections \cite{CHUNPO}.  Assuming no lepton flavour violation in
other sectors of the theory (e.g. in the case of supersymmetry) the
matrix $I$ is diagonal, $I_{AB}=I_A\delta_{AB}$ (the index $A$ refers
to the electroweak basis). 
Without loss of generality in our discussion we take $I_\mu = 0$.
We adopt the general form for the CP conserving mixing matrix
$U~$~\cite{Schechter:1980gr,PDG00}
\begin{eqnarray}
\left(\matrix{c_{12}c_{13}&s_{12}c_{13}&s_{13}\cr
-s_{12}c_{23}-c_{12}s_{23}s_{13}&c_{12}c_{23}-s_{12}s_{23}s_{13}&
s_{23}c_{13}\cr s_{12}s_{23}-c_{12}c_{23}s_{13}&
-c_{12}s_{23}-s_{12}c_{23}s_{13}&c_{23}c_{13}}\right)
\label{eqn:ckm}
\end{eqnarray}
in which $\theta_{13} \equiv \theta_{reactor}$, and allow for
different CP parities of neutrino eigenstates. 
For degenerate masses $|m_a|=m$ the matrix (\ref{eqn:ckm}) contains
one redundant angle and this degree of freedom is used to
re-diagonalize the mass matrix (\ref{eqn:massmatcor}).
As long as we assume CP conservation we can use the standard
perturbation theory for degenerate zeroth order eigenvalues in order
to determine the non-trivial relation amongst the three mixing angles
in \eq{eqn:ckm}.
Taking $|I_\tau| \gg |I_e|$ we easily recover the conclusions of ref.
\cite{CHKRPO}. In this letter we investigate the alternative
possibility that $|I_e| \gg |I_\tau|$.  We begin with
$m_1=-m_2=m_3\equiv m$ in which case
\begin{eqnarray}
m_\nu^{ab}=m\left(\matrix{1+2U^2_{A1}I_A & 0 & 2U_{A1}U_{A3}I_A\cr
                                 0&-1-2U^2_{A2}I_A&0\cr
                                 2U_{A1}U_{A3}I_A & 0 & 1+2U^2_{A3}I_A}\right)
\end{eqnarray}
Since $m_1=m_3$, the ordinary perturbation calculus tells us that the
neutrino mass basis should be chosen so that the off-diagonal $13$
entry of the perturbation is zero. We therefore require that
$U_{A1}U_{A3}I_A=0$ which fixes the redundant angle in the matrix
(\ref{eqn:ckm}). With our assumptions about $I_A$'s ($|I_e| \gg
|I_\tau|$) this gives
\begin{eqnarray}
s_{13}=-{s_{12}\over c_{12}}s_{23}c_{23}r+{\cal O}(r^2)
\end{eqnarray}
where $r\equiv I_\tau/I_e$.
It is then trivial to compute the corrected mass eigenvalues and one finds
\begin{eqnarray}
&&\Delta m^2_{atm} \approx -4m^2I_e s^2_{12}~, \label{eqn:dmatm}\\
&&\Delta m^2_\odot \approx -4m^2I_e\left[
C_{12}
\left(1-s^2_{23}r\right)
+  (1 + 2 c^2_{12}) s^2_{13}\right]
\label{eqn:dmsol}
\end{eqnarray}
where $C_{12} \equiv c^2_{12}-s^2_{12}$.  For maximal $\theta_{12} =
\pi/4$ one obtains that $\Delta m^2_{atm}\sim I_e$ and $\Delta
m^2_\odot\sim I_e r^2$, in agreement with the experimental requirement
$\Delta m^2_\odot\ll\Delta m^2_{atm}$.

Eqs. (\ref{eqn:dmatm}) and (\ref{eqn:dmsol}) lead to the following
relation between the neutrino oscillation parameters
\be{ddsol}
\Delta m^2_\odot\approx 2\Delta m^2_{atm}\left[
C_{12}
\left(1-s^2_{23}r\right)
+ 2 s^2_{13}\right]
\label{eqn:dmsdma}
\ee
which is the prediction of our model, independent of the value of
$I_e$.  Since we need $c_{12}^2\approx s_{12}^2$ we conclude that
small mixing solution to the solar neutrino problem is ruled out in
our neutrino unification picture.

It is easy to check that with our assumptions about $I_A$'s the same
mechanism works in the case $-m_1=m_2=m_3$. However, in the case
$m_1=m_2=-m_3$ the condition $U_{A1}U_{A2}I_A=0$ means that
$s_{12}c_{12}c^2_{13}={\cal O}(r)$ which means that one cannot have
simultaneously large solar neutrino mixing and remain consistent with
the reactor experiments~\cite{chooz} (this is also incompatible with
the SMA solution to the solar neutrino problem because for
$s_{12}c_{12}\approx0$ one gets $\Delta m^2_\odot\sim\Delta
m^2_{atm}$). 

% This case is also disfavored, for large $m$ values by
% neutrinoless double beta decay.

Since $\Delta m^2_{atm} \approx 3 \times 10^{-3}$eV$^2$ is rather
fixed by the analysis of the data~\cite{latestglobalanalysis} and
since, as we shall see, typically $|I_e| \lsim 10^{-3}$, it follows
that $m \gsim$ \O(1) eV. Thus this picture leads to neutrino masses in
the range accessible to $\beta$ decay and hot dark matter searches.
Different CP parities of the neutrinos are however sufficient to
ensure a destructive interference in the amplitude for neutrinoless
double beta decay which is suppressed by the fact that the two leading
mass eigenstate neutrinos in the \ne form a pseudo-Dirac
neutrino~\cite{Wolfenstein:1981rk}.

How can one realize our neutrino unification scenario?  First of all,
it requires the existence of new states giving a dominant contribution
to $I_e$.  Rather than extending the gauge and/or multiplet structure
of the SM, we prefer to ask whether in supersymmetry one can arrange
the corrections $I_A$ to satisfy our requirements.  Taking a diagonal
slepton mass matrix (in the same basis in which the charged leptons
mass matrix is diagonal) and taking into account only the contribution
of pure Wino of mass $\tilde m$ one finds for $I_A \equiv I (y_A)$ the
following form
\begin{eqnarray}
I_A  = {g^2\over 32 \pi^2}
\left \{ - {1\over y_A}+{y_A^2-1\over y_A^2}\ln (1-y_A)
- (A \to \mu) \right\} 
\end{eqnarray}
where $y_A=1-(M_A/\tilde m)^2$, $M_A$ is the $A$-th charged slepton
mass. It is easy to see that for $I_e \sim 10^{-3}$ and positive we
need $M_e \approx 1.7 M_{\mu, \tau}$ for the charged slepton masses.

Such slepton mass patterns can arise in models with inverted
hierarchy~\cite{DUPOSA}, although the proposed mechanism may have
other realizations.  Being an operator of dimension five, the
nontrivial neutrino mass matrix does not affect the evolution of the
lepton Yukawa and slepton mass squared matrices. Hence, if there
exists a basis at $M_X$ in which both lepton Yukawa and slepton mass
squared matrices are simultaneously diagonal (fermion-sfermion mass
alignment~\cite{NISE} present e.g. in models with abelian $U(1)$ gauge
symmetry~\cite{align}), this basis remains unchanged during the RG
evolution to the electroweak scale and hence the slepton masses remain
diagonal. Thus in this case lepton flavor-violating decays such as
$\mu \to e \gamma$ are not induced.

\begin{figure*}
  \epsfxsize30pc
\includegraphics[width=0.9\textwidth,height=4.5cm]{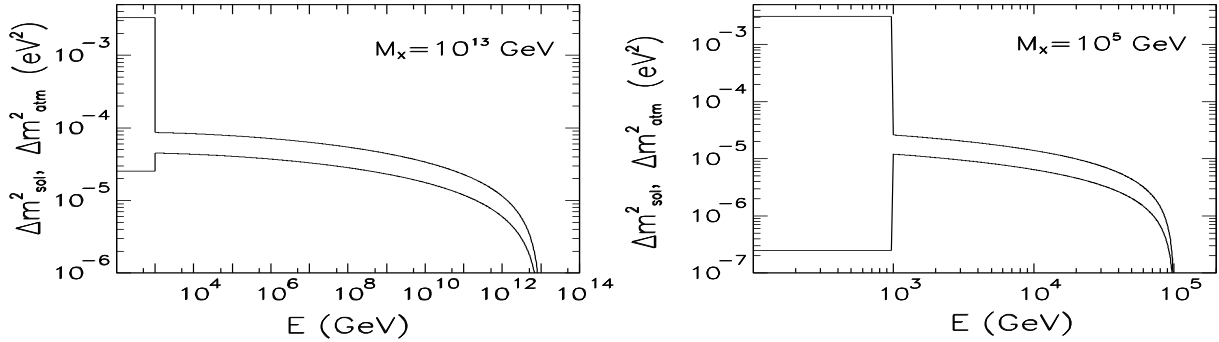} \hfil
\caption{Schematic evolution of solar and atmospheric 
  splittings corresponding to LMA (left) and LOW solutions (right)
  \label{fig}}
\end{figure*}

In \fig{fig} we illustrate the neutrino unification idea by displaying
the supersymmetric evolution of solar and atmospheric neutrino mass
splittings as a function of energy, from the high energy scale $M_X$
where neutrino masses unify, down to the weak scale for tan$\beta = 2$
and a starting neutrino mass at $M_X$ of 1.2 eV.
Above the supersymmetry breaking scale the neutrino mass-square
differences are $\Delta m^2_{21}= {1 \over 3} \Delta m^2_{31}= {1
  \over 2} \Delta m^2_{32} \simeq -m^2 I_\tau$. After threshold
corrections they change according to Eqs.
(\ref{eqn:dmatm},\ref{eqn:dmsol},\ref{eqn:dmsdma}). Depending on the
chosen value of $M_X$ this can fit large mixing angle solutions to the
solar neutrino problem.
For example for $M_X= 10^{13}$ GeV and $I_e \approx 2 \times 10^{-3}$,
$r=\frac{I_\tau}{I_e} = -0.029$ we obtain quasi-degenerate neutrino
masses $|m_i| = 0.9$ eV split by $\Delta m^2_{21}=-1.52 \times
10^{-5}$ eV$^2$ and $\Delta m^2_{32}=-3.13 \times 10^{-3}$ eV$^2$.
Neutrino mixing is bi-maximal $s_{12} \approx s_{23} \approx 0.707$
with $s_{13} \approx 0.0143$, nicely matching the LMA solution.
In order to achieve the LOW solution one needs a smaller $I_\tau$
value.  Barring a possible cancellation between the terms in the r.h.s
of Eq.  (\ref{eqn:dmsdma}) this requires a low neutrino unification
scale $M_X= 10^5$ GeV. In this case we find that quasi-degenerate
masses $|m_i| = 1.1$ eV for $r=\frac{I_\tau}{I_e} = -0.0103$ and $I_e
\approx 1.3 \times 10^{-3}$ with $\Delta m^2_{21}=-2.74 \times
10^{-7}$ eV$^2$ and $\Delta m^2_{32}=-3.16 \times 10^{-3}$ eV$^2$.
Mixing is  nearly bi-maximal with $s_{13} \approx 0.0051$.
Due to the observed flatness of the latest recoil electron energy
spectrum, present solar neutrino data globally prefer large mixing
angle MSW solutions over the small mixing solution (SMA)
~\cite{latestglobalanalysis}.
This feature fits well in our neutrino unification scheme, where the
SMA solution can {\sl not} be realized.
From ref.~\cite{latestglobalanalysis} we find that in our scheme the
LMA solution is present at the 97$\%$ C.L. while the LOW solution
appears already at 90$\%$ C.L.

\vskip0.2cm 
\noindent{\bf  Conclusions.}

We have proposed a simple theoretical ansatz where
neutrino masses arise from a common dimension-5 operator at some
``neutrino unification'' scale $M_X$. This scale must be lower than
that which characterizes the unification of gauge and Yukawa
couplings.  The ansatz naturally implies quasi-degenerate neutrinos at
the electron-volt range, thus accessible to future $\beta$ decay and
hot dark matter searches.  However neutrinoless double beta decay is
naturally suppressed due to the intrinsic Majorana neutrino CP
parities.
We have shown that neutrino splittings can be induced by
supersymmetric thresholds arising from non-universal soft breaking
terms, without necessarily conflicting with limits on lepton flavour
violating processes such as $\mu \to e \gamma$.  A non-trivial
consequence of the neutrino unification idea is that the solar and
atmospheric data can only be accounted for in terms of large mixing
angle-type MSW oscillations.
Note also that the consistency of our scheme correlates the smallness
of $\theta_{13}$ indicated by reactor data and, to a lesser extent,
also by the atmospheric data~\cite{latestglobalanalysis}, to the
largeness of $\theta_{12}$ required by the solar neutrino data.

\vskip 0.2cm

\noindent{\bf Acknowledgements:}  Work
supported by Spanish DGICYT under grants PB98-0693 and SAB/1998-0136
(A.I.), by the European Commission TMR network HPRN-CT-2000-00148, by
the European Science Foundation network grant N.  86 and by the Polish
State Committee for Scientific Research grant KBN 2 P03B 060 18.
P.H.Ch. thanks the CERN Theory group for hospitality during the
completion of this paper.

\end{document}